\documentclass{article}[fullpage]

\usepackage{amsfonts,amsmath,amsthm,amssymb,braket,enumerate,mathtools,xcolor,float}
\usepackage[margin=2cm]{geometry}
\usepackage[backref]{hyperref}
\usepackage{comment}
\usepackage{todonotes}
\usepackage{authblk}

\usepackage{tikz,ifdraft,wrapfig}
\usetikzlibrary{quantikz}
\usetikzlibrary{arrows,positioning,matrix,backgrounds,calc,fit}
\tikzcdset{
every matrix/.style={ampersand replacement=\&}
}

\usepackage{thmtools}
\usepackage{thm-restate}

\makeatletter
\renewcommand*\env@matrix[1][*\c@MaxMatrixCols c]{%
  \hskip -\arraycolsep
  \let\@ifnextchar\new@ifnextchar
  \array{#1}}
\makeatother

\newcommand{\A}{\mathbb{A}}
\newcommand{\C}{\mathbb{C}}
\newcommand{\I}{\mathbb{I}}
\newcommand{\N}{\mathbb{N}}
\newcommand{\Q}{\mathbb{Q}}
\newcommand{\R}{\mathbb{R}}

\newcommand{\Z}{\mathbb{Z}}
\newcommand{\cC}{\mathcal{C}}
\newcommand{\cD}{\mathcal{D}}

\newcommand{\cL}{\mathcal{L}}

\newcommand{\cU}{\mathcal{U}}
\newcommand{\K}{K}
\newcommand{\zd}{(\Z_d)}
\newcommand{\cij}{\gamma_{ij}}

\newcommand{\Tr}{T}
\newcommand{\TrE}{T^{E}}
\newcommand{\TrF}{T^{F}}
\newcommand{\tTrE}{T^{E}}
\newcommand{\tTrF}{T^{F}}

\newcommand{\Sr}{S}
\newcommand{\SrE}{S^{E}}
\newcommand{\SrF}{S^{F}}
\newcommand{\tSrE}{S^{E}}
\newcommand{\tSrF}{S^{F}}

\newcommand{\kz}{\ket{\vec{z\,}}}

\newcommand{\Spec}{\text{Spec}}

\newtheorem{theorem}{Theorem}
\newtheorem{defn}{Definition}

\newtheorem{lemma}{Lemma}

\theoremstyle{remark}
\newtheorem{remark}{Remark}
\newtheorem{example}[equation]{Example}

\author{Imin Chen}
\author{Nadish de Silva}

\affil{Department of Mathematics, Simon Fraser University, Burnaby, B.C., Canada}

\title{\vspace{-1.3cm}Characterising semi-Clifford gates using algebraic sets}

\begin{document}

\date{} 
\maketitle

\vspace{-0.7cm}\begin{abstract}
Motivated by their central role in fault-tolerant quantum computation, we study the sets of gates of the third-level of the Clifford hierarchy and their distinguished subsets of `nearly diagonal' semi-Clifford gates.  The Clifford hierarchy gates can be implemented via gate teleportation given appropriate magic states.  The vast quantity of these resource states required for achieving fault-tolerance is a significant bottleneck for the practical realisation of universal quantum computers.  Semi-Clifford gates are important because they can be implemented with far more efficient use of these resource states.  

We prove that every third-level gate of up to two qudits is semi-Clifford.  We thus generalise results of Zeng-Chen-Chuang (2008) in the qubit case and of the second author (2020) in the qutrit case to the case of qudits of arbitrary prime dimension $d$.  

Earlier results relied on exhaustive computations whereas our present work leverages tools of algebraic geometry. Specifically, we construct two schemes corresponding to the sets of third-level Clifford hierarchy gates and third-level semi-Clifford gates.  We then show that the two algebraic sets resulting from reducing these schemes modulo $d$ share the same set of rational points.
\end{abstract}

\vspace{-0.4cm}\tableofcontents

\newpage

\section{Introduction}
In this article, we are concerned with two classes of fundamental quantum computational operations.  The first contains those which are potentially costly to perform but are a standard ingredient in achieving universal quantum computation.  The second is a subset of the first: those which can be performed using a protocol that is significantly more efficient than the standard one.  We prove that, in common scenarios, these sets coincide.  Doing so leads to an application of tools and language of algebraic geometry to quantum information---including some which have yet to be applied to this field.  

In quantum computation, elementary units of information are represented by \emph{states}: vectors in $\C^{d}$ where $d \in \N$ is a number of discrete degrees of freedom of a physical medium.  The $d=2$ case corresponds to a \emph{qubit}, the quantum analogue of a classical bit.  While this is the most common case considered within quantum information, recently, more attention has been paid to the higher-dimensional \emph{qudit} case, e.g.\ the qutrit case of $d=3$.  Qudit-based computation promises increased capacity and efficiency \cite{quditmsd,tonchev2016quantum}.  The advantages of qudit-based computation has led to rapidly accelerating development by experimentalists \cite{chi2022programmable,chizzini2022molecular,karacsony2023efficient,low2020practical,ringbauer2022universal,seifert2022time,wang2018proof}.  In the future, once quantum technology has progressed, we expect qudit-based computation to become commonplace.  A system of $n$ qudits is modeled by  $(\C^{d})^{\otimes n} \simeq \C^{d^n}$.

\emph{Quantum gates}, the elementary physical and computational operations acting on these data, are represented by unitary operators on $\C^{d^n}$.  A universal quantum computer must be capable of applying any unitary transformation, to an arbitrarily good approximation, to its input data.  However, in practice, only a subgroup of the full unitary group is directly physically implemented.  This restriction is necessary for enabling the techniques of error correction and fault-tolerance required for building a large-scale practical quantum computer.

The most common family of quantum error-correcting codes are stabiliser codes \cite{gottthesis}.  In these schemes, data are encoded as eigenvectors of \emph{Pauli gates}.  Members of the group of \emph{Clifford gates} 
are special in that they can often be easily applied fault-tolerantly to data protected by stabiliser codes.  Quantum universality, however, further requires the ability to fault-tolerantly perform non-Clifford gates.  The standard choices of non-Clifford gates come from a set called the \emph{third level of the Clifford hierarchy}.  These can be fault-tolerantly performed via the \emph{gate teleportation} protocol using only Clifford gates supplemented with ancillary \emph{magic state} resources \cite{gottesman1999}.


A significant practical barrier to achieving quantum universality via the supplementation of Clifford gates with magic states is the need to prepare such states for every desired application of a non-Clifford gate.  The original gate teleportation protocol implemented $n$-qubit third-level gates using magic states of $2n$ qubits.   The need to reduce the burden of this substantial resource overhead cost led to the study of more efficient gate teleportation protocols.

Third-level qubit gates that are diagonal in the standard basis can be implemented using magic states of only $n$ qubits \cite{zhou2000methodology}.  This more efficient gate teleportation protocol was generalised to allow performing `nearly diagonal' \emph{semi-Clifford gates}, i.e.\ those gates $G$ such that  $G = C_1 D C_2$ for $C_1,C_2$ being Clifford gates and $D$ a diagonal gate of the Clifford hierarchy \cite{zeng2008semi}.  
Research into the Clifford hierarchy and semi-Clifford gates has remained active from their discovery twenty-five years ago to the present \cite{anderson2022groups,cui2017diagonal,oldpaper,gottesman1999,pllaha2020weyl,rengaswamy2019unifying,zeng2008semi,zhou2000methodology}.

Zeng-Chen-Chuang \cite{zeng2008semi} proved that all Clifford hierarchy gates of one or two qubits are semi-Clifford; as they restricted their results to the single case of $d=2$, they were able to employ a proof involving exhaustive computations.  Beigi-Shor \cite{beigi2009c3} showed that, other than the case of three-qubit third-level gates, this no longer holds in the setting of more than two qubits.  

More recent work of the second author \cite{oldpaper} initiated the study of qudit semi-Clifford gates via a unifying perspective based on the finite-dimensional Stone-von Neumann theorem.  The efficient gate teleportation protocols for qubit semi-Clifford gates were generalised to the qudit case, ensuring that the notion of semi-Clifford gate is still of interest in the qudit setting.  In this work, it was proved that all third-level gates of one qudit or two qutrits are semi-Clifford.  In the case of two qutrits, the proof relied on exhaustive computations.

Below, we generalise the existing results for two qubits or qutrits to the case of arbitrary odd prime dimension, thus establishing that all third-level gates of up to two qudits are semi-Clifford.

\begin{restatable}[Main theorem]{theorem}{mainthm}
\label{mainthm}
For any odd prime dimension $d \in \N$, every two-qudit third-level gate $G \in \cC^2_3$ is semi-Clifford.
\end{restatable}
  

\subsection{Mathematical methods}

In order to establish our result for infinitely many dimensions simultaneously, our mathematical arguments are necessarily highly abstract.  We anticipate that our methods (and their theoretical justification) are applicable more widely within quantum information and beyond.

We first transform our original question into one about two systems of polynomial equations over the finite field $\Z_d$.  That is, we show that is sufficient to prove our theorem for a subset of two-qudit third-level gates that can be described by solutions to a family $\mathcal{F}_1$ of polynomials over $\Z_d$.  Next, we show that those solutions of $\mathcal{F}_1$ that satisfy an additional family  $\mathcal{F}_2$ must describe semi-Clifford third-level gates.  It is therefore sufficient to establish that the radical ideals of the the generated ideals $(\mathcal{F}_1), (\mathcal{F}_1 \cup \mathcal{F}_2) \subseteq \Z_d[x_1,...x_n]$ coincide.

In theory, this can be verified computationally using Gr\"obner bases.  In practice, the polynomial systems $\mathcal{F}_1, \mathcal{F}_2$ may be too complex to practically compute.  This is the case in our problem and we supply the necessary methods for simplifying them.  One technique we employ is to rewrite a system of polynomial equations as an inhomogeneous linear system over the field of fractions of a polynomial ring. This allows us to consider a weakened polynomial system which results from imposing that this linear system is consistent, thereby reducing the number of variables and rendering our computations feasible. 

The analysis of our simplified equations naturally benefits from the application of geometric intuition.  Although the weakened polynomial system introduces extraneous components to its space of solutions, we are able to show that these components are disjoint with the original polynomial system on rational points.  Our result follows from showing that two geometric spaces (specifically, \emph{algebraic sets}), one arising from the set of third-level gates and the other from its subset of semi-Clifford gates, share the same set of (rational) points.  

Using \emph{schemes}, an even more abstract notion of geometric space, we are to able establish our result for all odd prime dimensions with only one series of computations on polynomial systems with coefficients in $\Z[1/2]$.

\section{Preliminaries}

To render this article accessible to both mathematicians and quantum information theorists, we provide background information on quantum computing and algebraic geometry.

\subsection{Quantum computation}

Let $d \in \N$ be an odd prime and $n \in \N$.  Denote by $\omega = e^{i 2 \pi / d}$ the $d$-th primitive root of unity and by $[n]$ the set $\{1,...,n\}$.  The \emph{computational basis} of $(\C^d)^{\otimes n}$ is the standard basis whose members are denoted by the kets $\kz$ for $\vec{z\,} \in \Z_d^n$ where $\Z_d$ is the ring of integers modulo $d$.  

The unitary group of $(\C^d)^{\otimes n}$ is denoted $\cU(d^n)$.  When we refer to a unitary $U$ \emph{up to phase} we mean its equivalence class $[U]$ under the equivalence relation $U \sim V \iff U = e^{i\theta}V$ for some $\theta \in \R$.  Given $n$ unitaries $U_1, ..., U_n$ and a vector $\vec{p}$ of $n$ integers, we denote by $U^{\vec{p}}$ the product $U_1^{p_1} \cdots U_n^{p_n}$.

\subsubsection{Pauli gates}

The Pauli gates form the basis of the stabiliser codes of quantum error correction in that quantum data are encoded as simultaneous eigenvectors of commuting sets of Pauli gates.  

The \emph{basic Pauli gates} for a single qudit are $Z \in \cU(d)$ and $X \in \cU(d)$: $Z \ket{z} = \omega^{z} \ket{z}$ and $X \ket{z} = \ket{z +1}$ where the addition is taken modulo $d$; these unitaries have order $d$.  For $n>1$ qudits and $i \in [n]$, define $Z_i \in \cU(d^n)$ to be a tensor product of $n-1$ identity matrices of size $d \times d$ with $Z$ in the $i$-th factor: $\I \otimes ...  \otimes  Z \otimes ...  \otimes  \I$; $X_i$ is defined similarly.  The $Z_i, X_i$ are the \emph{$n$-qudit basic Pauli gates} and they satisfy the Weyl commutation relations: \begin{equation}Z_i X_i = \omega X_i Z_i\end{equation}  and, for $i \neq j$, the pairs $Z_i,X_j$; $Z_i,Z_j$; and, $X_i,X_j$ commute.  

For each $\vec{p},\vec{q\,} \in \Z_d^n$ we define the following unitaries as products of the basic Pauli gates: 
\begin{equation}Z^{\vec p} = Z_1^{p_1} \cdots Z_n^{p_n} \quad \quad X^{\vec q} = X_1^{q_1} \cdots X_n^{q_n}.\end{equation}

\begin{defn}\label{paulidefn}The group of \emph{Pauli gates} is the subgroup of $\cU(d^n)$ generated by the basic Pauli gates: $$\cC^n_1 = \{\omega^c Z^{\vec{p}} X^{\vec{q\,}}\ \; | \; c \in \Z_d, (\vec{p}, \vec{q\,}) \in \Z_d^{2n} \}.$$
\end{defn}



The set of pairs $(\vec{p},\vec{q\,}) \in \Z_d^{2n}$ form a symplectic vector space over $\Z_d$ with the symplectic product: 
\begin{equation}
\label{sympprod}
    [(\vec{p},\vec{q}),(\vec{p\,}',\vec{q\,}')] = \vec{p} \cdot \vec{q\,}' - \vec{p\,}' \cdot \vec{q}.
\end{equation}  
Pauli gates obey the commutation relation: 
\begin{equation}
\label{commpaulisymp}
    Z^{\vec{p}}X^{\vec{q}} \cdot Z^{\vec{p\,}'}X^{\vec{q\,}'} = \omega^{[(\vec{p},\vec{q}),(\vec{p\,}',\vec{q\,}')]}   Z^{\vec{p\,}'}X^{\vec{q\,}'} \cdot Z^{\vec{p}}X^{\vec{q}}.
\end{equation}
\begin{remark}\label{reindexz4}
    It will be more convenient below, when studying two-qudit gates ($n = 2$), to reindex the coordinates of $(\vec{p},\vec{q\,}) = ({p_1}, {p_2}, {q_1}, {q_2}) \in \Z_d^{4}$ as $({p_1}, {q_1}, {p_2}, {q_2})$.  This ordering is aligned with that of Definition \ref{conjtupdef} which collects the pairs of conjugates of $Z_i, X_i$ for $i \in [n]$.  The symplectic product of Equation \eqref{sympprod} remains the same: \begin{equation}[({p_1}, {q_1}, {p_2}, {q_2}),({p}'_1, {q}'_1, {p}'_2, {q}'_2)] = {p_1}q'_1 + {p_2}q'_2 - {p'_1}q_1 - {p'_2}q_2.\end{equation}
\end{remark}

\subsubsection{Conjugate tuples}

If we conjugate $Z,X \in \cU(d)$ by any unitary gate $G$, the gates $G_Z = GZG^*$ and $G_X = GXG^*$ also satisfy the Weyl commutation relations: ${G_Z}^d = {G_X}^d = \I$ and $G_ZG_X = \omega G_XG_Z$.  Remarkably, if $(U,V)$ is any pair of unitaries satisfying these relations, there exists a gate $G$, unique up to phase, such that $U = G_Z$ and $V = G_X$.  

\begin{defn}
An ordered pair of unitaries $(U,V) \in \cU(d^n) \times \cU(d^n)$ is a \emph{conjugate pair} if
\begin{enumerate}\itemsep0em 
\item $U^d = \I$ and $V^d = \I$,
\item $UV = \omega VU$.
\end{enumerate}
\end{defn}

There is a bijective correspondence between one-qudit gates $G \in \cU(d)$ (up to phase) and conjugate pairs $(U,V)$.  This is a consequence of the Stone-von Neumann theorem \cite[Lemma 3.4]{oldpaper}.  This notion naturally extends to gates of $n$ qudits. 

\begin{defn}\label{conjtupdef}
A \emph{conjugate tuple} $( (U_i,V_i) )_{i \in [n]}$ is an ordered tuple of $n$ conjugate pairs of $n$-qudit gates such that any two elements of distinct pairs commute.
\end{defn}

There is a bijective correspondence between $n$-qudit gates $G \in \cU(d^n)$ (up to phase) and conjugate tuples \cite[Lemma 3.8]{oldpaper}.

\begin{defn}
The \emph{conjugate tuple of an $n$-qudit gate} $G\in \cU(d^n)$ is $( (GZ_iG^*,GX_iG^*) )_{i \in [n]}$.
\end{defn}

Below, we study gates via their conjugate tuples.


\subsubsection{Clifford gates}

Clifford gates often have a simple fault-tolerant implementation on stabiliser codes.  

\begin{defn}\label{cliffdefn}The group of \emph{Clifford gates} is the normaliser of the group of Pauli gates as a subgroup of $\cU(d^n)$: \begin{equation}\cC_2^n = \{C \in \cU(d^n) \;|\; C \cC_1^n C^* \subseteq \cC_1^n \}.\end{equation}
\end{defn}

\begin{lemma}\label{commcliff}
For any Clifford gate $C \in \cC^n_2$ and Pauli gate $P_1 \in \cC^n_1$, there exists a Pauli gate $P_2 \in \cC^n_1$ such that \begin{equation}CP_1 = P_2C.\end{equation}
\end{lemma}

\begin{defn}A gate $G_1 \in \cU(d^n)$ is \emph{Clifford-conjugate} to $G_2 \in \cU(d^n)$ if there exists a Clifford gate $C \in \cC^n_2$ such that $G_1 = CG_2C^*$.
\end{defn}

\noindent 

The Clifford gates, up to phase, are in correspondence with affine symplectic transformations of $\Z_d^{2n}$.  First, we define the group of  Clifford gates up to phase: $[\cC_2^n] = \cC_2^n / \mathbb{T}$, where $\mathbb{T}$ is the complex unit circle group.  The group $Sp(n,\Z_d) \ltimes \Z_d^{2n}$ of affine symplectic transformations of $\Z_d^{2n}$ are pairings of $2n \times 2n$ symplectic matrices over $\Z_d$ and translations in $\Z_d^{2n}$ with the composition law: \begin{equation}(A,v) \circ (B,w) = (AB, Aw + v).\end{equation}  There is a (\!\emph{Weil} or \emph{metaplectic}) projective representation $\rho: Sp(n,\Z_d) \ltimes \Z_d^{2n} \to [\cC_2^n]$ that is an isomorphism between the groups of affine symplectic transformations and Clifford gates up to phase \cite{gross2006hudson, neuhauser2002explicit}.  This representation extends to the case of finite fields of odd characteristic which is more general than the case treated in this work.



The group $\cC_2^n$ is a maximal nondense subgroup of $\cU(d^n)$, and thus, approximately performing an arbitrary computation requires the ability to fault-tolerantly perform a non-Clifford gate.

\subsubsection{Third-level gates}

The standard choices for non-Clifford gates to supplement the group of Clifford gates in order to achieve universal quantum computation come from the \emph{Clifford hierarchy}.  This is a recursively-defined and nested sequence of subsets of $\cU(d^n)$.  The groups of Pauli and Clifford gates form the first and second \emph{levels} respectively: $\cC_1^n \subset \cC_2^n \subset \cC_3^n \subset ...$.  

Non-Clifford gates that are in the third level or higher can be implemented fault-tolerantly on encoded data indirectly via a gate teleportation protocol.  Such a protocol takes as input an arbitrary data state and a resource magic state and as output produces, using only Clifford gates and standard measurements, the data state with the desired gate applied to it.  We are concerned in this work only with third-level gates as these can be implemented with the simplest gate teleportation protocols and are, as such, the standard choices in practical quantum computational schemes.  

The set of third-level gates is defined very similarly to how the group of Clifford gates is defined.  It is the set of gates that conjugate any Pauli gate to yield a Clifford gate.  The simplest qudit example is the $T$ gate \cite{howard2012qudit} defined by Howard-Vala.

\begin{defn}\label{thirdlvldefn}The set of \emph{third-level gates} is the subset of $\cU(d^n)$: \begin{equation}\cC_3^n = \{G \in \cU(d^n) \;|\; G \cC_1^n G^* \subseteq \cC_2^n \}.\end{equation}
\end{defn}

Note that $\cC_3^n$ is closed under multiplication on the left and right by elements of $\cC_2^n$.

From the above definition, we see that the conjugate tuple of a third-level gate consists of Clifford gates.  Conversely, under the correspondence between gates (up to phase) and conjugate tuples, any conjugate tuple of Clifford gates yields a third-level gate.  This was first observed in the qubit case by Beigi-Shor \cite{beigi2009c3} and extended to higher dimensions and levels of the Clifford hierarchy in \cite{oldpaper}.

\begin{theorem}[{\cite[Theorem 3.12]{oldpaper}}]\label{C3fromtuples}
Gates of the third-level of the Clifford hierarchy, up to phase, are in bijective correspondence with conjugate tuples of Clifford gates.
\end{theorem}


Of particular interest is the subset of the Clifford hierarchy gates which are diagonal in the computational basis.

\begin{defn}\label{diagkdefn}The group of \emph{diagonal $k$-th-level gates} is: \begin{equation}\cD_k^n = \{G \in \cC_k^n \;|\; G \text{ is a diagonal gate}\}.\end{equation}
\end{defn}

\subsubsection{Semi-Clifford gates}

Zhou-Leung-Chuang \cite{zhou2000methodology} introduced a simplified gate teleportation protocol, based on Bennett-Gottesman's one-qubit teleportation, capable of fault-tolerantly implementing certain qubit Clifford hierarchy gates using half the ancillary resources required in the original Gottesman-Chuang protocol.  This class of gates includes the diagonal Clifford hierarchy gates.  Zeng-Chen-Chuang \cite{zeng2008semi} introduced the notion of semi-Clifford gates which are `nearly diagonal' in the sense of being within Clifford corrections of diagonal gates:

\begin{defn}\label{sc}A third-level gate $G \in \cC_3^n$ is \emph{semi-Clifford} if $G = C_1 D C_2$ where $C_1,C_2 \in \cC_2^n$ are Clifford gates and $D \in \cD_3^n$ is a diagonal third-level gate.\end{defn}

The following gate teleportation protocol for implementing the qudit semi-Clifford gate $G = C_1 D C_2$ using the magic state $\ket{M} = D\ket{+}$ was introduced in \cite[\S5(a)]{oldpaper}.  It ensures that the notion of semi-Clifford is still relevant in the qudit setting.

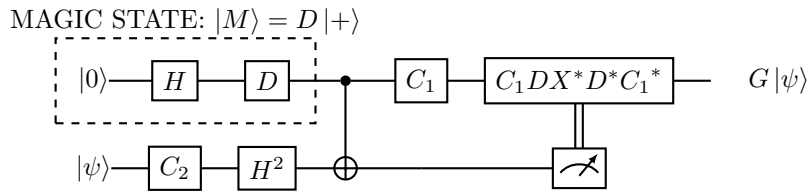
\begin{center}
\begin{quantikz}
\ket{0}\gategroup[wires=1,steps=3, style={dashed}]{{MAGIC STATE: $\ket{M} = D\ket{+}$}}    \& \gate{H} \& \gate{D} \& \ctrl{1} \& \gate{C_1}   \& \gate{C_1 D X^* D^* {C_1}^*} \& \qw \& G\ket{\psi} \\
\ket{\psi} \& \gate{C_2} \& \gate{H^2} \& \targ{} \& \qw    \& \meter{}  \vcw{-1}
\end{quantikz}
\end{center}

The following two lemmas are a direct consequence of Definition~\ref{sc}.

\begin{lemma}\label{cliffmultsemi}If $G \in \cC_3^n$ is semi-Clifford then, for any Clifford $C \in \cC^n_2$,  $GC$ and $CG$ are also semi-Clifford.
\end{lemma}

\begin{lemma}\label{cliffconjsemi}If $G_1 \in \cU(d^n)$ is Clifford-conjugate to $G_2 \in \cU(d^n)$, then $G_1$ is semi-Clifford if and only if $G_2$ is.
\end{lemma}

In \cite{oldpaper}, an equivalent characterisation is given for the property of a gate being semi-Clifford.

\begin{defn}\label{lagsemi}A \emph{Lagrangian semibasis} of a symplectic vector space of dimension $2n$ is a linearly independent set of $n$ vectors $\{v_1, ..., v_n\}$ satisfying $[v_i, v_j] = 0$ for all $i,j \in [n]$.
\end{defn}

Motivating the definition of a Langrangian semibasis is the fact that there exists a Clifford gate $C \in C^n_2$ such that $CZ_iC^* = \omega^{c_i} Z^{\vec p_i}X^{\vec q_i}$ for $i \in [n], c_i \in \Z_d, \vec p_i,\vec q_i \in \Z^n_d$ if and only if $\{(\vec p_i,\vec q_i)\}_{i \in [n]}$ is a Lagrangian semibasis.

\begin{theorem}[{\cite[Theorem 5.4]{oldpaper}}]\label{semicliff}
Suppose $G \in \cC_k^n$ and denote its conjugate tuple by $( (U_i, V_i) )_{i \in [n]}$.  $G$ is semi-Clifford if and only if there exists a Lagrangian semibasis $\{(\vec{p_i}, \vec{q_i})\}_{i \in [n]} \subseteq \Z^{2n}_d$ such that, for each $i \in [n]$, $U^{\vec{p_i}} V^{\vec{q_i}}$ is a Pauli gate.
\end{theorem}

\subsubsection{Quadratic and almost diagonal Clifford gates}

Here we define two special classes of Clifford gates that will arise as members of the conjugate tuples of third-level gates below.

For any $n \times n$ symmetric matrix $\Phi$ over $\Z_d$ the $2n \times 2n$ block matrix \begin{equation}S = \begin{pmatrix}
\I & 0 \\
\Phi & \I \end{pmatrix}\end{equation} is symplectic with respect to the symplectic pairing of Equation \eqref{sympprod} above.  Under the representation  $\rho$ of \cite{neuhauser2002explicit}, the image of $(S,0) \in Sp(n,\Z_d) \ltimes \Z_d^{2n}$ is, up to phase, the diagonal Clifford gate $D_\Phi \in \cD_2^n$ defined by $D_\Phi\kz = \omega^{{\vec z \,}^t \Phi \vec z} \kz$.

\begin{defn}\label{quadcliff}A \emph{quadratic Clifford gate} $D_\Phi \in \cD_2^n$ is a diagonal gate of the form $D_\Phi\kz = \omega^{{\vec z \,}^t \Phi \vec z} \kz$ where $\Phi$ is a $n \times n$ symmetric matrix over $\Z_d$.
\end{defn}

\begin{defn}
A Clifford gate $C \in \cC_2^n$ is \emph{almost diagonal} if it is of the form $e^{i \theta} D_\Phi P$ where $\theta \in \R$, $D_\Phi \in \cD_2^n$ is a quadratic Clifford gate, and $P \in \cC^n_1$ is a Pauli gate.
\end{defn}

The almost diagonal Clifford gates form a group; we will require only the following lemma.

\begin{lemma}\label{proddiagcliff}
    If $C_1 = e^{i \theta_1}D_{\Phi_1} P_1 \in \cC_2^n$ and $C_2 = e^{i \theta_2} D_{\Phi_2} P_2 \in \cC_2^n$ are two almost diagonal Clifford gates, then $C_1C_2$ is the almost diagonal Clifford gate $e^{i (\theta_1 + \theta_2)} D_{\Phi_1 + \Phi_2} P$ for some Pauli gate $P \in \cC^n_1$.
\end{lemma}

\begin{proof}\begin{align}    C_1C_2 &=  e^{i (\theta_1 + \theta_2)} D_{\Phi_1} P_1 \cdot D_{\Phi_2} P_2 \\ 
    &= e^{i (\theta_1 + \theta_2)} D_{\Phi_1} D_{\Phi_2} P_3 P_2\\
    &= e^{i (\theta_1 + \theta_2)} D_{\Phi_1 + \Phi_2} P    \end{align}

The second equality follows from Lemma \ref{commcliff}.  The last equality follows choosing $P = P_3 P_2$ and because \begin{equation}D_{\Phi_1} D_{\Phi_2} \kz = \omega^{{\vec z \,}^t \Phi_1 \vec z} \omega^{{\vec z \,}^t \Phi_2 \vec z} \kz = \omega^{{\vec z \,}^t \Phi_1 \vec z + {\vec z \,}^t \Phi_2 \vec z} \kz = \omega^{{\vec z \,}^t (\Phi_1 + \Phi_2) \vec z} \kz = D_{\Phi_1 + \Phi_2} \kz.\end{equation}   \end{proof}

\subsubsection{Characterisation of simplified third-level gates via a polynomial system}

In this subsection, we define a subset of the third-level gates that are easily described by variables in $\Z_d$.

\begin{defn}
A third-level gate $G \in \cC_3^n$ is \emph{simplified} if its conjugate tuple contains only almost diagonal Clifford gates.
\end{defn}

By using the symplectic formalism for Clifford gates and applying a classification of the maximal abelian subgroups of the symplectic group over finite fields due to Barry \cite{barry1979large}, we can `simplify' any third-level gate. 

\begin{theorem}[{\cite[Lemma 5.7]{oldpaper}}]\label{diagsymp}
Every third-level gate $G \in \cC_3^n$ is Clifford-conjugate to a simplified one.
\end{theorem}

There is a map from simplified two-qudit third-level gates $G$ to the set of solutions to a certain system of 18 polynomial equations over $\Z_d$ in 28 variables.  As we shall see, such a third-level gate is semi-Clifford if and only if its image under this map satisfies additional polynomial constraints.

Let $G \in \cC^2_2$ be a simplified two-qudit third-level gate and denote its conjugate tuple by $( (U_1,V_1),(U_2,V_2) )$, i.e.\ \begin{equation}U_1 = GZ_1G^*, \quad V_1 = GX_1G^*, \quad U_2 = GZ_2G^*, \quad V_2 = GX_2G^*.\end{equation}

Since $G$ is simplified, $U_1 = e^{i \theta_1} D_{\Phi_1} P_1$ is a quadratic Clifford gate.  Expressing $P_1$ as $\omega^{c_1} Z^{\vec p_1} Z^{\vec q_1}$, we see that $U_1$ is characterised up to phase by $\Phi_1$ (a $2 \times 2$ symmetric matrix over $\Z_d$), $\vec p_1 \in \Z_d^2,$ and $ \vec q_1 \in \Z_d^2$: that is, by 7 elements of $\Z_d$.  

We similarly define $(\Phi_i, \vec p_i, \vec q_i)$ for $i \in \{2,3,4\}$ to characterise $V_1,U_2,V_2$ up to phase respectively.  We can give necessary conditions for four septuples of $\Z_d$ to arise in this way from a conjugate tuple of a simplified third-level gate.  The following theorem follows directly from the definition of conjugate tuple and \cite[Lemma 5.8]{oldpaper}.

\begin{theorem}\label{polyeqns}
The ordered pair of ordered pairs of order $d$ almost diagonal Clifford gates \begin{equation}\label{ctupleadcg}( (e^{i \theta_1} D_{\Phi_1}Z^{\vec p_1}X^{\vec q_1},\,e^{i \theta_2} D_{\Phi_2}Z^{\vec p_2}X^{\vec q_2}),\;(e^{i \theta_3} D_{\Phi_3}Z^{\vec p_3}X^{\vec q_3},\,e^{i \theta_4} D_{\Phi_4}Z^{\vec p_4}X^{\vec q_4}) )\end{equation} is the conjugate tuple of a simplified two-qudit third-level gate only if its variables satisfy the following set of polynomial equations for $1 \leq i < j \leq 4$:

\begin{align}
    \label{first-equ}\tag{T1} & \Phi_i \vec q_j = \Phi_j \vec q_i \\
    \label{second-equ}\tag{T2} & \vec q_i^{\;t} \Phi_j \vec q_i - \vec q_j^{\;t} \Phi_i \vec q_j + \vec p_i \cdot \vec q_j - \vec p_j \cdot \vec q_i = \cij 
\end{align}
where
\begin{equation}\label{third-equ}
  \cij = \begin{cases}
    1 & \text{ if } (i,j) \in \left\{ (1,2), (3,4) \right\} \\
    0 & \text{ otherwise}.
    \end{cases}
\end{equation}
\end{theorem}

\begin{remark}
    The Equations \eqref{first-equ}, \eqref{second-equ} are necessary for the almost diagonal Clifford gates of Equation \eqref{ctupleadcg} to obey the commutation relations of a conjugate tuple.  As these relations are unaffected by phase, $e^{i\theta_1}, \dots, e^{i\theta_4}$ can be ignored.
\end{remark}

We can also give necessary and sufficient conditions for a simplified two-qudit third-level gate to be semi-Clifford in terms of the variables that characterise the quadratic Clifford parts in its conjugate tuple.

\begin{theorem}\label{kernel}
Let $G \in \cC^2_3$ be a simplified two-qudit third-level gate with conjugate tuple \begin{equation}( (U_1 = e^{i \theta_1} D_{\Phi_1}P_1,\,V_1 = e^{i \theta_2} D_{\Phi_2}P_2),\;(U_2 = e^{i \theta_3} D_{\Phi_3}P_3,\,V_2 = e^{i \theta_4} D_{\Phi_4}P_4) ).\end{equation}   Express the symmetric matrices $\Phi_i$ as 
\begin{equation}\begin{pmatrix}
\Phi_{i1} & \frac{1}{2}\Phi_{i3} \\
 \frac{1}{2}\Phi_{i3} & \Phi_{i2} \\\end{pmatrix}.\end{equation} Then $G$ is semi-Clifford if and only if the kernel of the matrix
\begin{equation}\begin{pmatrix}
\Phi_{11} & \Phi_{21} & \Phi_{31} & \Phi_{41} \\
\Phi_{12} & \Phi_{22} & \Phi_{32} & \Phi_{42} \\
\Phi_{13} & \Phi_{23} & \Phi_{33} & \Phi_{43} \end{pmatrix}\end{equation} contains a Lagrangian semibasis of $\Z_d^4$.
\end{theorem}

\begin{proof}
    For any column vector $(\rho_1, \kappa_1, \rho_2, \kappa_1) \in \Z_d^4$, which is indexed as in Remark \ref{reindexz4}, the unitary $U^{\vec \rho} V^{\vec \kappa}$ satisfies \begin{align}
        e^{-i (\rho_1 \theta_1 + \kappa_1 \theta_2 + \rho_2 \theta_3 + \rho_2 \theta_4)} U^{\vec \rho} V^{\vec \kappa} &= (D_{\Phi_1}P_1)^{\rho_1} \cdots (D_{\Phi_4}P_4)^{\kappa_2}\\
        &= (D_{\Phi_1})^{\rho_1} \cdots (D_{\Phi_4})^{\kappa_2} \cdot P \\
        &= D_{\rho_1 \Phi_1 + \kappa_1 \Phi_2 + \rho_2 \Phi_3 + \kappa_2 \Phi_4} \cdot P
    \end{align}      
    for some Pauli gate $P \in \cC^2_1$ by repeated application of Lemma \ref{proddiagcliff}.  An almost diagonal Clifford gate $D_{\Phi} P$ is a Pauli gate if and only if $\Phi$ is the zero matrix.   Thus, $U^{\vec \rho} V^{\vec \kappa}$ is a Pauli gate if and only if $(\rho_1, \kappa_1, \rho_2, \kappa_1)$ is in the kernel of the above matrix and the result follows from Theorem \ref{semicliff}.
\end{proof}



\subsection{Algebraic geometry}\label{algeo}

Here, we briefly review the elements of algebraic geometry necessary for understanding our proof.  First, we recall the notion of \emph{algebraic sets}, which are geometric objects arising as the solution set to polynomial families, the correspondence between these sets and certain ideals in a polynomial ring, and their decomposition into irreducible components.  We then briefly review \emph{schemes}, a generalisation of algebraic sets that admits reduction modulo primes; this allows us to perform one series of computations to prove our result for all qudit dimensions. 

\subsubsection{Algebraic sets}

\label{algebraic-sets}

Let $k$ be a field and $R = k[x_1, \ldots, x_n]$ the polynomial ring in variables $x_1, \ldots, x_n$ with coefficients in $k$. A nonempty subset $I \subseteq R$ is an ideal if it satisfies:
\begin{enumerate}
    \item for all $a, b \in I$, $a - b \in I$,
    \item for all $a \in I$ and $r \in R$, $r a \in I$.
\end{enumerate}

\begin{example}
Let $f_1, \ldots, f_m \in R$. Then
\begin{equation}
    I = R f_1 + \ldots + R f_m
\end{equation}
is an ideal of $R$, called the ideal generated by $f_1, \ldots, f_m$ and denoted by $(f_1, \ldots, f_m)$.
\end{example}

We first describe the dictionary of classical algebraic geometry in order to orient the reader.  It is not necessary to replace $k$ by an algebraic closure of $k$ in order for the theory to work, but it is technically easier to understand and provides intuition for the more general theory of schemes that we will use to describe our computations.

We define \emph{affine $n$-space} to be the set $\A^n = \K^n$ where $\K$ is a choice of algebraic closure of $k$. An \emph{algebraic set} in $\A^n$ is the \emph{vanishing set} of an ideal of $\K[x_1, \ldots, x_n]$, i.e.\ a subset of the form:
\begin{equation}
    V(I) = \left\{ (x_1, \ldots, x_n) \in \A^n \;|\; \forall f \in I \; f(x_1, \ldots, x_n) = 0   \right\}.
\end{equation}
For $B$ being any subset of $R$, we define
\begin{equation}
     V(B) = V(\langle B \rangle)
\end{equation}
where $\langle B \rangle$ is the intersection of all ideals of $R$ containing $B$, called the ideal generated by $B$.
\begin{remark}
    An algebraic set in $\A^n$ is just the solution set of a subset of polynomials in $\K[x_1, \ldots, x_n]$.
\end{remark}

\begin{theorem} (Hilbert basis)
Let $I$ be an ideal of $R$. Then $I = (f_1, \ldots, f_m)$ for a finite number of $f_1, \ldots, f_m \in R$.
\end{theorem}

The \emph{radical} of an ideal $I$ in $R$ is the ideal in $R$ given by
\begin{equation}
    \sqrt{I} = \left\{ f \in R \;|\; \exists k \in \N \; f^k \in I \right\},
\end{equation}
and we say an ideal $I$ in $R$ is a radical ideal if $\sqrt{I} = I$. The radical of an ideal $I$ in $R$ is the smallest radical ideal containing $I$.

An ideal $I$ in $R$ is \emph{prime} if $fg \in I$ implies $f \in I$ or $g \in I$. A prime ideal is automatically a radical ideal from these definitions. 
\begin{theorem} (Nullstellensatz)
\label{nullstellensatz}
There is a bijective correspondence between algebraic sets in $\A^n$ and radical ideals in $\K[x_1, \ldots, x_n]$.
\end{theorem}

Under this correspondence, a radical ideal $I \subset \K[x_1, \ldots, x_n]$ is mapped to the algebraic set $V(I) \subset \A^n$ and an algebraic set $S \subset \A^n$ is mapped to the radical ideal: \begin{equation}I(S) = \{f \in \K[x_1, \ldots, x_n] \;|\; \forall (x_1,...,x_n) \in S \; f(x_1,...,x_n) = 0\}.\end{equation}

The \emph{Zariski topology} on $\A^n$ is given by taking the algebraic sets to be the collection of closed sets. We say an algebraic set $S$ is \emph{irreducible} if $S$ is not a union of two proper closed subsets of $\A^n$. In terms of the Nullstellensatz correspondence, irreducible algebraic sets correspond to prime ideals of $\K[x_1, \ldots, x_n]$: $I(S)$ is prime whenever $S$ is an irreducible algebraic set. Irreducible algebraic sets in $\A^n$ are called (affine) varieties.

Every algebraic set $S$ can be uniquely (up to order) decomposed into a finite number of irreducible algebraic sets $S_1, \ldots, S_m$, called the \emph{irreducible components} of $S$, in the sense that
\begin{equation}
\label{irred-decomp}
    S  = S_1 \cup \ldots \cup S_m,
\end{equation}
where
\begin{equation}
    S_i \not \subseteq \cup_{j \not= i} S_j.
\end{equation} 
By Theorem~\ref{nullstellensatz}, the decomposition in \eqref{irred-decomp} translates  into the so-called \emph{radical decomposition} of the ideal $I$
\begin{equation}
\label{radical-decomposition}
    I = I_1 \cap \ldots \cap I_m,
\end{equation}
where $I$ is the radical ideal of $\K[x_1, \ldots, x_n]$ corresponding to $S$ and $I_1, \ldots, I_m$ are the prime ideals in $\K[x_1, \ldots, x_n]$ corresponding to the irreducible components $S_1, \ldots, S_m$, respectively, i.e.\ $I = I(S)$ and $I_i = I(S_i)$.

We now describe a more arithmetic version of this theory which is intermediate between classical algebraic geometry and the general theory of schemes and often used in explicit arithmetic geometry.

An algebraic set $S \subseteq \A^n$ is \emph{defined over $k$} if its corresponding ideal $I(S)$ in $\K[x_1, \ldots, x_n]$ is generated by polynomials in $k[x_1, \ldots, x_n]$. If we define
\begin{equation}
    I(S/k) = I(S) \cap R
\end{equation}
then an algebraic set $S \subseteq \A^n$ is defined over $k$ if and only if
\begin{equation}
    I(S) = I(S/k) \K[x_1, \ldots, x_n]
\end{equation}
(see \cite{Silverman}). The \emph{$k$-rational points} of an algebraic set $S$ defined over $k$ is the set 
\begin{equation}
    S(k) = \left\{ (x_1, \ldots, x_n) \in k^n \;|\, \forall f \in I(S/k) \; f(x_1, \ldots, x_n) = 0\right\}. 
\end{equation}

If $S \subseteq \A^n$ is defined over $k$ and $I \subseteq k[x_1, \ldots, x_n]$ is its corresponding ideal, there is a version of \eqref{radical-decomposition} in $R$ which asserts that
\begin{equation}
\label{radical-decomposition-rational}
    I = I_1' \cap \ldots \cap I_{m'}',
\end{equation}
where the $I_i'$ are prime ideals in $R$. Geometrically, this corresponds to a decomposition
\begin{equation}
    S = S_1' \cup \ldots \cup S_{m'}'
\end{equation}
where the $S_i'$ are algebraic sets defined over $k$ which are `irreducible' over $k$ (in a sense made precise in the next section). Furthermore we have that
\begin{equation}
\label{rational-point-decomposition}
    S(k) = S_1'(k) \cup \ldots \cup S_{m'}'(k).
\end{equation}

We will not prove \eqref{radical-decomposition-rational} $\implies$ \eqref{rational-point-decomposition} as it is a special case of a more general decomposition proven in the next section. Also note that, in this formulation, commutative algebra is used to prove the corresponding geometric statement about rational points rather than the other direction in \eqref{irred-decomp} $\implies$ \eqref{radical-decomposition}.

\subsubsection{The language of schemes}

The notion of an algebraic set $S \subset \A^n$ defined over $k$ and its `irreducibility' over $k$ is best described using Grothendieck's theory of schemes. The theory is quite general and allows consideration of the analogues of algebraic sets over a commutative ring $A$.  While a thorough introduction to the theory of schemes is beyond the scope of this article, we review the elements required for our proof and refer the reader to \cite{Eisenbud-Harris} for a more complete treatment.

The principal construction in this theory is to associate to a commutative ring $R$ a topological space: the set
\begin{equation}
    \Spec(R) = \left\{ \mathfrak{p} : \mathfrak{p} \text{ a prime ideal of } R \right\},
\end{equation}
under the Zariski topology where closed subsets are given by
\begin{equation}
  V(I) = \left\{ \mathfrak{p} \in \Spec(R) : \mathfrak{p} \supseteq I \right\}.
\end{equation}
It is endowed with a sheaf of rings (coming from localisations of $R$) so that it becomes a locally ringed space called the \emph{affine scheme of $R$}. In this construction, ring homomorphisms $\varphi: R \rightarrow S$ induce and correspond to morphisms $\varphi^* : \Spec(S) \rightarrow \Spec(R)$ by mapping a prime $\mathfrak{q}$ of $\Spec(S)$ to the prime ideal $\varphi^{-1}(\mathfrak{q})$ of $\Spec(R)$.

A closed subscheme of $\Spec(R)$ is given by $\Spec(R/I)$ for an ideal $I$ of $R$ and comes with a morphism $\Spec(R/I) \rightarrow \Spec(R)$.  This morphism gives a homeomorphism of $\Spec(R/I)$ with the closed subset $V(I)$ of $\Spec(R)$.

More generally, a \emph{scheme} $S$ is a locally ringed space possessing an open cover with each set in the cover being isomorphic to an affine scheme.

If $R$ is an $A$-algebra, where $A$ is a commutative ring, then there is a structure morphism 
\begin{equation}
    \Spec(R) \rightarrow \Spec(A),
\end{equation}
and we say $\Spec(R)$ is a \emph{scheme over $\Spec(A)$} (often this is abbreviated as scheme over $A$). For example, when $R = k[x_1, \ldots, x_n]/I$ with $I$ an ideal of $k[x_1, \ldots, x_n]$ and $A = k$, there is a structure morphism
\begin{equation}
\label{structure-morphism}
    \Spec(k[x_1, \ldots, x_n]/I) \rightarrow \Spec(k),
\end{equation}
corresponding to the inclusion
\begin{equation}
    k \hookrightarrow k[x_1, \ldots, x_n]/I,
\end{equation}
where we note $R$ then has the structure of a commutative $k$-algebra. We now define the analogue of a solution contained within $k^n$: a  \emph{$k$-rational point} on $\Spec(R)$ (also called a \emph{$k$-section}), is a morphism 
\begin{equation}
\label{section}
    \Spec(k) \rightarrow \Spec(R)
\end{equation}
that upon composition with the structure morphism in \eqref{structure-morphism} yields the identity morphism on $\Spec(k)$. In terms of rings, it means that a $k$-rational point of $\Spec(R)$ corresponds to a $k$-algebra homomorphism $R \rightarrow k$.

In the case that $k$ is algebraically closed and $R = k[x_1, \ldots, x_n]/I$ is the $k$-algebra that is the quotient by an ideal $I$ of $k[x_1, \ldots, x_n]$, the $k$-rational points of $\Spec(R)$ are in bijection with points of the algebraic set $V(I) \subseteq k^n$. More generally, if $S$ is a scheme over a commutative ring $A$, we denote by $S(A)$ the set of $A$-sections (also called the set of $A$-rational points) of $S$ defined as morphisms 
\begin{equation*}
    \Spec(A) \rightarrow \Spec(R)
\end{equation*}
that upon composition with the structure morphism yields the identity on $\Spec(A)$.

In case that $S$ is a scheme over a commutative ring $A$ and we have a ring homomorphism $A \rightarrow A_0$ to another commutative ring $A_0$, it is understood that $S(A_0)$ means  $(S \times_{A} A_0)(A_0)$ where $S \times_{A} A_0$ is the `base change' of $S$ from $A$ to $A_0$ described in the next paragraph.

If $S = \Spec(R)$ is an affine scheme, then the \emph{base change} $S \times_{A} A_0$ is the scheme $\Spec(R \otimes_{A} A_0)$ over $A_0$, where $R \otimes_{A} A_0$ is the tensor product of $R$ and $A_0$ as $A$-modules.  We can define the base change of a general scheme $S$ from $A$ to $A_0$ by using an open cover of $S$ consisting of affine schemes and patching together the base changes of these open affine schemes. This is a special case of a more general notion of fiber product of two schemes with morphisms to a base scheme \cite{Eisenbud-Harris}.

\subsubsection{A decomposition of schemes over $\Z[1/2]$}

\label{decomposition}

We now prove a refinement of \eqref{radical-decomposition-rational}, \eqref{rational-point-decomposition}.
\begin{theorem}
\label{scheme-decomposition}
Let $R$ be a commutative ring which is an $A$-algebra for a commutative ring $A$ and suppose $I, I_1, \ldots, I_m$ are ideals in $R$ such that
\begin{align}
   I = I_1 \cap \ldots \cap I_m.
\end{align}
Let $S = \Spec(R/I), S_1 = \Spec(R/I_1), \ldots, S_m = \Spec(R/I_m)$ be the corresponding schemes over $A$. Suppose we have a ring homomorphism $A \rightarrow A_0$ where $A_0$ is a field. Then
\begin{align}
\label{mod-d-rational-points}
   S(A_0) = S_1(A_0) \cup \ldots \cup S_m(A_0).
\end{align}
\end{theorem}
\begin{proof}
An $A_0$-rational point in $S(A_0)$ corresponds to a maximal ideal $\mathfrak{m}$ of the $A$-algebra $R$. If $\mathfrak{m} \supseteq I$, then it follows that $\mathfrak{m} \supseteq I = I_1 \cap \ldots \cap I_{m} \supseteq I_1 \cdots I_m$. Since $\mathfrak{m}$ is a maximal ideal, it is a prime ideal, and hence we have that $\mathfrak{m} \supseteq I_j$ for some $j = 1, \ldots, m$, that is, the $A_0$-rational point occurs in one of the sets $S_j(A_0)$.

An $A_0$-rational point in $S_j(A_0)$ corresponds to a maximal ideal $\mathfrak{m}$ of the $A$-algebra $R$. Since $\mathfrak{m} \supseteq I_j$, we have that $\mathfrak{m} \supseteq I = I_1 \cap \ldots \cap I_m$ and hence the $A_0$-rational point lies in $S(A_0)$.
\end{proof}
In the language of schemes, the above says that $S$ is a scheme over $A$ and decomposes into a union of closed subschemes $S_1, \ldots, S_m$ over $A$; hence, there is a corresponding decomposition of the set of its $A_0$-rational points. The union $S_1 \cup \ldots \cup S_n$ is the closed subscheme over $A$ given by the ideal $I_1 \cap \ldots \cap I_m$. Let $\iota(S), \iota_1(S_1), \ldots, \iota_m(S_m)$ be the closed subsets of $\Spec(R)$ associated to the closed subschemes $S, S_1, \ldots, S_m$. Following the same proof as Theorem~\ref{scheme-decomposition}, we have that $\iota(S) = \iota_1(S_1) \cup \ldots \cup \iota_m(S_m)$.

\begin{remark}
\label{general-decomposition}
In the above theorem, the ideals $I, I_1, \ldots, I_m$ need not be prime nor radical.
\end{remark}

In our application of interest, $A = \Z[1/2]$ and $A_0 = \Z_d$ for $d \not= 2$. Theorem~\ref{scheme-decomposition} gives a method to establish \eqref{mod-d-rational-points} for all $d \not= 2$ with one computation which we now describe. 

Suppose we have a scheme $X$ over $\Z$ given by an ideal $I \subseteq \Z[x_1, \ldots, x_n]$, i.e.\ $X = \Spec(\Z[x_1, \ldots, x_n]/I)$. We first compute a probable radical decomposition of $I \otimes \Q$ over $\Q$ using {\tt Magma} using the built-in function {\tt ProbableRadicalDecomposition}. This corresponds to 
\begin{align}
   I \otimes \Q = I_1' \cap \ldots \cap I_m',
\end{align}
for some ideals $I_1', \ldots, I_m' \subseteq \Q[x_1, \ldots, x_n]$.

\begin{remark}
In order for the above computation in {\tt Magma} to complete, we often have to use the probable version of radical decomposition. This means that the components returned may still decompose further, i.e.\ they may not correspond to prime ideals. However, this does not affect the arguments because of Remark~\ref{general-decomposition}.
\end{remark}

Let $I_1, \ldots, I_m$ be the ideals in $\Z[x_1, \ldots, x_n]$ generated by generators of $I_1', \ldots, I_m'$, respectively, but scaled to lie in $\Z[x_1, \ldots, x_n]$ by removing the greatest common factor among the coefficients. We next check that
\begin{align}
\label{tensor-half}
  I \otimes \Z[1/2] = (I_1 \otimes \Z[1/2]) \cap \ldots \cap (I_m \otimes \Z[1/2]),
\end{align}
using {\tt Magma} and the following lemma.
\begin{lemma}
\label{tensor-up}
If $I, I_1, \dots, I_m \subseteq \Z[x_1, \dots, x_n]$ are ideals then  \[ (I_1 \cap \ldots \cap I_m) \otimes \Z[1/2] =  (I_1 \otimes \Z[1/2]) \cap \ldots \cap (I_m \otimes \Z[1/2]). \]
\end{lemma}
\begin{proof}
   This follows from \cite[Theorem 7.4(i)]{Matsumura} as $\Z[1/2]$ is flat over $\Z$ .
\end{proof}
We achieve \eqref{tensor-half} by first computing the intersection
\begin{align}
  I_1 \cap \ldots \cap I_m,
\end{align}
in $\Z[x_1, \ldots, x_n]$ which is possible as {\tt Magma} is able to compute intersections and equality of ideals (using the {\tt meet} and {\tt eq} operators respectively) in $\Z[x_1, \ldots, x_n]$. Applying Lemma~\ref{tensor-up}, we obtain
\begin{align}
  (I_1 \cap \ldots \cap I_m) \otimes \Z[1/2] = (I_1 \otimes \Z[1/2]) \cap \ldots \cap (I_m \otimes \Z[1/2]).
\end{align}
We then verify using {\tt Magma} that
\begin{align}
\label{decomposition-half}
  (I \otimes \Z[1/2]) = (I_1 \otimes \Z[1/2]) \cap \ldots \cap (I_m \otimes \Z[1/2])
\end{align}
by checking that each generator of the LHS of \eqref{decomposition-half} is a $\Z[1/2][x_1, \ldots, x_n]$-linear combination of the generators of the RHS of \eqref{decomposition-half} and vice versa. If the verification succeeds, we can apply Theorem~\ref{scheme-decomposition} with $A = \Z[1/2]$ and $A_0 = \Z_d$ for $d \not= 2$ to conclude that
\begin{align}
  X(\Z_d) = X_1(\Z_d) \cup \ldots \cup X_m(\Z_d)
\end{align}
for all $d \not=2$,  where $X_j = \Spec(\Z[x_1, \ldots, x_n]/I_j)$ for $j = 1, \ldots, m$. Specifically in applying Theorem~\ref{scheme-decomposition}, we take
\begin{align}
    S & = \Spec(\Z[1/2][x_1, \ldots, x_n]/(I\otimes\Z[1/2]), \\
    S_j & = \Spec(\Z[1/2][x_1, \ldots, x_n]/(I_j \otimes \Z[1/2]), 
\end{align}
for $j = 1, \ldots, m$, and $S, S_j$ are the base changes of $X, X_j$ from $\Z$ to $\Z[1/2]$, respectively.

\section{Main Result}\label{mainresult}

\subsection{Overview}

\mainthm*

Let $d \in \N$ be an odd prime dimension and let $G \in \cC^2_3$ be a two-qudit third-level gate.  By Theorem \ref{diagsymp}, $G$ is Clifford-conjugate to a simplified third-level gate and, so, by Lemma \ref{cliffconjsemi}, we can assume without loss of generality that $G$ is a simplified third-level gate.

Therefore, the conjugate tuple $( (U_1,V_1),(U_2,V_2) )$ of $G$ consists of almost diagonal gates: \begin{align*}
U_1 &= GZ_1G^* \,= e^{i\theta_1} D_{\Phi_1}Z^{\vec p_1}X^{\vec q_1}\\
V_1 &= GX_1G^* = e^{i\theta_2} D_{\Phi_2}Z^{\vec p_2}X^{\vec q_2}\\
U_2 &= GZ_2G^* \,= e^{i\theta_3} D_{\Phi_3}Z^{\vec p_3}X^{\vec q_3}\\
V_2 &= GX_2G^* = e^{i\theta_4} D_{\Phi_4}Z^{\vec p_4}X^{\vec q_4}\end{align*}
where $\theta_i \in \R$, $\vec p_i, \vec q_i \in \Z^2_d$, and $\Phi_i$ is a $2 \times 2$ symmetric matrix over $\Z_d$:$$\Phi_i  = \begin{pmatrix}
      \Phi_{i1} & \frac{1}{2} \Phi_{i3} \\
      \frac{1}{2} \Phi_{i3} & \Phi_{i2} 
    \end{pmatrix} .$$

Recall that, by Theorem \ref{polyeqns}, for all $1 \le i < j \le 4$:
\begin{align*}
    & \Phi_i \vec q_j = \Phi_j \vec q_i \tag{\ref{first-equ}}\\
    & \vec q_i^{\;t} \Phi_j \vec q_i - \vec q_j^{\;t} \Phi_i \vec q_j + \vec p_i \cdot \vec q_j - \vec p_j \cdot \vec q_i = \cij \tag{\ref{second-equ}}
\end{align*}
where
\begin{equation*}
  \cij = \begin{cases}
    1 & \text{ if } (i,j) \in \left\{ (1,2), (3,4) \right\} \tag{\ref{third-equ}}\\
    0 & \text{ otherwise}.
    \end{cases}
\end{equation*}

Ultimately, our aim is to show that $G$ is semi-Clifford.  We will do this by showing that the solution in $\Z_d$ of Equations \eqref{first-equ}, \eqref{second-equ} yielded by $G$ satisfies an additional three polynomial equations (Equations \eqref{semicliffpoly}) that guarantee $G$ is semi-Clifford (Lemma \ref{semibasis}).  As the number of variables and the complexity of the systems of polynomial equations involved render the necessary calculations infeasible, we make two simplifications.  

First, we show that without loss of generality, we may assume that $GZ_1G^*$ is a Pauli gate or, equivalently, that the matrix $\Phi_1 = 0$ (Theorem \ref{phireduction}).  Aside from eliminating three variables, making this reduction simplifies the polynomial conditions  of Equations \eqref{semicliffpoly} that ensure $G$ is semi-Clifford.

Our last reduction is to replace Equations \eqref{second-equ} with a pair of polynomials $E,F$ such that satisfying Equations \eqref{second-equ} implies that either $E = 0$ or $F = 0$ (Lemma \ref{reducetolinear}).  The immediate effect of this will be to eliminate eight more variables: the $\vec{p}_i$.  As a consequence of weakening some of our original constraints, we enlarge the corresponding spaces of solutions.  However, we will find that, of the solutions to either of the two new polynomial systems, those that arise from simplified third-level gates also satisfy the semi-Clifford equations.

\subsubsection{Algebraic sets and schemes}

For a fixed value of the dimension $d$, the above strategy is easily summarised in the language of algebraic sets.  We will define a set denoted by $\Tr\zd$ ($T$ for \emph{third-level}) of $\Z_d$-rational points: the set of $\Z_d$-solutions of Equations \eqref{first-equ}, \eqref{second-equ} yielded by simplified third-level gates $G$ with the reduction $\Phi_1 = 0$.   We will also define a subset denoted by $\Sr\zd$ ($S$ for \emph{semi-Clifford}): the subset of $\Tr\zd$ corresponding to solutions that also satisfy the semi-Clifford equations.  Thus, we aim to prove that $\Tr\zd = \Sr\zd$.

In theory, this can be checked computationally from the polynomial equations involved.  Each algebraic set over $\Z_d$ corresponds to an ideal in a polynomial ring with coefficients in $\Z_d$. To show two algebraic sets over $\Z_d$ have the same $\Z_d$-rational points it suffices to show the two corresponding ideals have the same radical ideal since the Nullstellensatz would imply that the two algebraic sets are the same.  However, as alluded to above, the size and complexity of the polynomial systems involved means that computationally checking the two ideals have the same radical is not feasible.

The motivates our final reduction that decreases the number of variables at the expense of enlarging the algebraic set $\Tr\zd$ to a union of two algebraic sets: $\TrE\zd \cup \TrF\zd$.  Here, the superscripts $E$ (resp. $F$) indicates that the $\Z_d$-rational points satisfy $E =0$ (resp. $F=0$) rather than Equations \eqref{second-equ}.  Let $\SrE\zd$ (resp. $\SrF\zd$) be the algebraic subset of $\TrE\zd$ (resp. $\TrF\zd$) of solutions that also satisfy the polynomial conditions to be semi-Clifford.  It now becomes computationally feasible to show $\TrE\zd$ decomposes into $5$ algebraic subsets $\TrE_1\zd, \ldots, \TrE_5\zd$ and $\SrE\zd$ decomposes into $3$ algebraic subsets $\SrE_1\zd, \ldots, \SrE_3\zd$ and the $\TrE_i\zd = \SrE_i\zd$ for $i = 1, 2, 3$, but $\TrE_4\zd, \TrE_5\zd$ are disjoint from $\Tr\zd$.  We establish the same facts replacing $F$ for $E$. These facts will imply that $\Tr\zd = \Sr\zd$.

Our ultimate goal is to prove  $\Tr\zd \subset \Sr\zd$ for \emph{all} odd prime dimensions $d$ (Lemma \ref{mainthmlemma}).  Using the above method to show $\Tr\zd = \Sr\zd$, a priori one would have to computationally decompose $\TrE\zd$ and $\SrE\zd$ separately for each odd prime $d$. To enable one computation that deals with all odd primes, we introduce in Section \ref{schemesoi} the corresponding schemes $\Tr, \Sr, \TrE, \SrE, \TrF, \SrF$ over $\Z[1/2]$.  In Section \ref{proofmain}, we decompose $\TrE$ into schemes $\TrE_1, \ldots \TrE_5$ over $\Z[1/2]$ and $\SrE$ into schemes $\SrE_1, \ldots, \SrE_3$ over $\Z[1/2]$; we have similar decompositions for $F$.  Reducing these decompositions modulo $d \not= 2$ gives the required decompositions to conclude $\Tr\zd = \Sr\zd$ as in the argument above.

\subsection{Reduction to an easier case}

Here, we show that we can assume without loss of generality that $GZ_1G^*$ is a Pauli gate.

\begin{theorem}\label{phireduction}
Let $G \in \cC^2_3$ be a simplified two-qudit third-level gate with conjugate tuple $$( (U_1 = e^{i \theta_1} D_{\Phi_1}P_1,\,V_1 = e^{i \theta_2} D_{\Phi_2}P_2),\;(U_2 = e^{i \theta_3} D_{\Phi_3}P_3,\,V_2 = e^{i \theta_4} D_{\Phi_4}P_4) ).$$  There exists a Clifford gate $C \in \cC^2_2$ such that $GC \in \cC^2_3$ is a simplified two-qudit third-level gate whose conjugate tuple is $$( (U'_1 = e^{i \theta'_1} D_{\Phi'_1}P'_1,\,V'_1 = e^{i \theta'_2} D_{\Phi'_2}P'_2),\;(U'_2 = e^{i \theta'_3} D_{\Phi'_3}P'_3,\,V'_2 = e^{i \theta'_4} D_{\Phi'_4}P'_4) )$$ with $\Phi'_1$ equal to the $2 \times 2$ zero matrix.
\end{theorem}

\begin{proof}
Since the matrices $\Phi_i$ are four members of the three-dimensional vector space over $\Z_d$ of symmetric $2 \times 2$ matrices, a nontrivial linear combination $\rho_1 \Phi_1 + \kappa_1 \Phi_2 + \rho_2 \Phi_3 + \kappa_2 \Phi_4 = 0$ holds.  

There exists $(\rho'_1,\kappa'_1,\rho'_2,\kappa'_2) \in \Z_d^4$, which is indexed as in Remark \ref{reindexz4}, such that $\{(\rho_1,\kappa_1,\rho_2,\kappa_2),(\rho'_1,\kappa'_1,\rho'_2,\kappa'_2)\}$ form a Lagrangian semibasis.  

To see this, note that of the $d^4$ elements of $\Z_d^4$, $d^3$ of these have vanishing symplectic product with $(\rho_1,\kappa_1,\rho_2,\kappa_2)$ while $d^4 - d$ of them are not a scalar multiple of $(\rho_1,\kappa_1,\rho_2,\kappa_2)$.  Since $d^3 > d$, $(d^4 - d) + d^3 > d^4$ and there must exist $(\rho'_1,\kappa'_1,\rho'_2,\kappa'_2) \in \Z_d^4$ satisfying both of these conditions.    Let $Q_1 = Z_1^{\rho_1} X_1^{\kappa_1} Z_2^{\rho_2} X_2^{\kappa_2}$ and $Q_2 = Z_1^{\rho'_1} X_1^{\kappa'_1} Z_2^{\rho'_2} X_2^{\kappa'_2}$.

The Pauli gates $Q_1, Q_2$ are independent (in the sense that no nontrivial product of them is a scalar multiple of the identity) since $\{(\rho_1,\kappa_1,\rho_2,\kappa_2),(\rho'_1,\kappa'_1,\rho'_2,\kappa'_2)\}$ is linearly independent and they commute using Equation \eqref{commpaulisymp} since $[(\rho_1,\kappa_1,\rho_2,\kappa_2),(\rho'_1,\kappa'_1,\rho'_2,\kappa'_2)] = 0$.  We can thus construct a Clifford gate $C \in \cC_2^2$ such that $C Z_1 C^* = Q_1$ and $C Z_2 C^* = Q_2$ by applying Lemma 5.3 of \cite{oldpaper}.  

Let $P \in \cC^2_1$ be a Pauli gate and let $CPC^* \in \cC^2_1$ be the Pauli gate $\omega^c Z^{\vec p}X^{\vec q}$ for $c \in \Z_d, \vec p, \vec q \in \Z_d^2$.  Then, 
\begin{align*}GC P C^*G^* &= \omega^c G(Z^{\vec p}X^{\vec q})G^*\\ 
&= \omega^c G(Z_1^{p_1})G^* \cdots G(X_2^{q_2})G^*\\ 
&= \omega^c {U}_1^{p_1} {U}_2^{p_2}{V}_1^{q_1}{V}_2^{q_2} \\
&= \omega^c (D_{\Phi_1}P_1)^{p_1} \cdots (D_{\Phi_4}P_4)^{q_2}\\
&= D_{p_1 \Phi_1 + q_1 \Phi_2 + p_2 \Phi_3 + q_2 \Phi_4} S
\end{align*} for some Pauli gate $S \in \cC^2_1$ by Lemma \ref{proddiagcliff}; the result is thus an almost diagonal Clifford gate.  By taking $P$ to be each of the basic Pauli gates, we see that $GC \in \cC^2_3$ is a simplified two-qudit third-level gate.

Taking $P = Z_1$ and noting that $CPC^* = Q_1 = \omega^0 Z^{\vec \rho}X^{\vec \kappa}$  we see that $GC Z_1 C^*G^* = GQ_1G^* = U^{\vec \rho} V^{\vec \kappa}$ which, following the preceding paragraph, has the form $D_{\Phi'_1}P'_1$ with $\Phi'_1 = \rho_1 \Phi_1 + \kappa_1 \Phi_2 + \rho_2 \Phi_3 + \kappa_2 \Phi_4 = 0$.
\end{proof}

\subsection{The semi-Clifford condition as a polynomial system}

We derive a set of polynomial constraints such that, under the reductions of Theorem \ref{phireduction}, solutions to Equations \eqref{first-equ}, \eqref{second-equ} that satisfy these additional constraints describe simplified third-level gates that are semi-Clifford.

We will employ the characterisation of two-qudit simplified semi-Clifford gates given by Theorem \ref{kernel}.  Recall that in the $n=2$ case, a Lagrangian semibasis (Definition \ref{lagsemi}) is a linearly independent pair of vectors with vanishing symplectic inner product.

\begin{lemma}
\label{semibasis}
If $\Phi_{11} = \Phi_{12} = \Phi_{13} = 0$, the kernel of the matrix
\begin{equation}
\label{kernel-matrix}
\begin{pmatrix}
\Phi_{11} & \Phi_{21} & \Phi_{31} & \Phi_{41} \\
\Phi_{12} & \Phi_{22} & \Phi_{32} & \Phi_{42} \\
\Phi_{13} & \Phi_{23} & \Phi_{33} & \Phi_{43} \end{pmatrix}
\end{equation}
contains a Lagrangian semibasis if and only if the following three equations are satisfied:
\begin{align} \label{semicliffpoly}
  \Phi_{31} \Phi_{42} - \Phi_{32} \Phi_{41} &= 0 \nonumber\\
  \Phi_{31} \Phi_{43} - \Phi_{33} \Phi_{41} &= 0 \tag{S}\\ 
  \Phi_{32} \Phi_{43} - \Phi_{33} \Phi_{42} &= 0. \nonumber 
\end{align}

\end{lemma}

\begin{proof}

Here, we consider the kernel as a subset of $\Z_d^4$, indexed as in Remark \ref{reindexz4} and with its corresponding symplectic product.  First, note that $(1,0,0,0)$ belongs to the kernel.  To form a Lagrangian semibasis from $(1,0,0,0)$, it is sufficient to have another nonzero vector of the form $(0, 0, k_3, k_4)$ in the kernel of \eqref{kernel-matrix}.  This is because  $[(1,0,0,0),(0, 0, k_3, k_4)] = 1\cdot 0 + 0\cdot k_4 - 0\cdot 0 - k_3\cdot 0 = 0$ and, if either $k_3 \not= 0 $ or $k_4 \not=0$, the two vectors form a linearly independent pair.  Thus, $\{(1,0,0,0),(0, 0, k_3, k_4)\}$ is a Lagrangian semibasis.

In fact, as we shall now prove, the existence of a nonzero vector $(0, 0, k_3, k_4)$ in the kernel is also necessary for the kernel to contain a Lagrangian semibasis.

Suppose that $\vec a = (a_1, a_2, a_3, a_4)$ and $\vec b = (b_1, b_2, b_3, b_4)$ form a Lagrangian semibasis in the kernel of the above matrix.  The span of $\vec a$ and $\vec b$ will also be inside the kernel and, by bilinearity of the symplectic product, any pair inside this span will also have vanishing symplectic product.

We can assume without loss of generality that $a_2$ is 0.  If $a_2 \neq 0$ and $b_2 = 0$, then we can swap the labels of $\vec a,\vec b$.  If both are nonzero then by replacing $\vec{a}$ with $a_2^{-1}\vec a - b_2^{-1}\vec b$, we have that $\{\vec a, \vec b\}$ is a Lagrangian semibasis with $a_2 = 1 - 1 = 0$.

We may further assume $a_1 = 1$.  If it is zero, then the lemma is proved by choosing the vector $(0,0,k_3,k_4)$ to be $(0,0,a_3,a_4)$.  If $a_1$ is not zero, then we can replace $\vec a$ with $a_1^{-1}\vec a$.  We have that $\{\vec a, \vec b\}$ is a Lagrangian semibasis with $\vec a = (1,0,a_3,a_4)$.

We know that $(1,0,0,0)$ is in the kernel and $[(1,0,0,0),(1, 0, a_3, a_4)] = 0$.  Thus $\{(1,0,0,0),\vec a - (1,0,0,0)\} = \{(1,0,0,0),(0,0,a_3,a_4)\}$ is a Lagrangian semibasis inside the kernel.  Choosing $k_3 = a_3$ and  $k_4 = a_4$, we have proved the kernel contains a nonzero vector $(0, 0, k_3, k_4)$.

Combining this with our first observation, we find that the kernel contains a Lagrangian semibasis if and only if it contains a nonzero vector $(0, 0, k_3, k_4)$.

Finally, we note that the condition that the kernel contains a nonzero vector $(0,0,k_3,k_4)$ is equivalent to the condition that the two rightmost columns are linearly dependent which is in turn is equivalent to all the $2 \times 2$ minors of the two rightmost columns vanishing.  The polynomial equations \eqref{semicliffpoly} capture these conditions. \end{proof}

\subsection{Weakening a polynomial system via linearisation}

Due to the number of variables and the complexity of the polynomial systems of Equations \eqref{first-equ}, \eqref{second-equ} that we use to describe simplified third-level gates, we must make some simplifications in order to render the necessary calculations feasible.

Here, we will weaken Equations \eqref{second-equ}: for all $1 \le i < j \le 4$
\begin{equation*}\vec q_i^{\;t} \Phi_j \vec q_i - \vec q_j^{\;t} \Phi_i \vec q_j + \vec p_i \cdot \vec q_j - \vec p_j \cdot \vec q_i = \cij \tag{\ref{second-equ}}\end{equation*}
where
\begin{equation*}
  \cij = \begin{cases}
    1 & \text{ if } (i,j) \in \left\{ (1,2), (3,4) \right\} \tag{\ref{third-equ}}\\
    0 & \text{ otherwise}.
    \end{cases}
\end{equation*}
We can rearrange \eqref{second-equ} by moving the quadratic terms to the right-hand side.  The result is an inhomogeneous linear system of equations.
\begin{equation}\tag{\ref{second-equ}'}
\label{linear-sys}
\begin{pmatrix}
    \vec q_2^{\;\, t} & - \vec q_1^{\;\, t} & 0 & 0 \\
    \vec q_3^{\;\, t} & 0 & - \vec q_1^{\;\, t} & 0 \\ 
    \vec q_4^{\;\, t} & 0 & 0 & - \vec q_1^{\;\, t}  \\ 
    0 & \vec q_3^{\;\, t} & - \vec q_2^{\;\, t} & 0  \\ 
    0 & \vec q_4^{\;\, t} & 0 & - \vec q_2^{\;\, t}  \\ 
    0 & 0 & \vec q_4^{\;\, t} & - \vec q_3^{\;\, t}  \\ 
\end{pmatrix}
\begin{pmatrix}
\vec p_1 \\
\vec p_2 \\
\vec p_3 \\
\vec p_4
\end{pmatrix} = 
\begin{pmatrix*}[l]
\,\vec q_2^{\;t} \,\Phi_1\, \vec q_2 - \vec q_1^{\;t} \,\Phi_2\, \vec q_1 + 1\,   \\
\,\vec q_3^{\;t} \,\Phi_1\, \vec q_3 - \vec q_1^{\;t} \,\Phi_3\, \vec q_1 \,\\
\,\vec q_4^{\;t} \,\Phi_1\, \vec q_4 - \vec q_1^{\;t} \,\Phi_4\, \vec q_1 \,\\
\,\vec q_3^{\;t} \,\Phi_2\, \vec q_3 - \vec q_2^{\;t} \,\Phi_3\, \vec q_2 \,\\
\,\vec q_4^{\;t} \,\Phi_2\, \vec q_4 - \vec q_2^{\;t} \,\Phi_4\, \vec q_2 \,\\
\,\vec q_4^{\;t} \,\Phi_3\, \vec q_4 - \vec q_3^{\;t} \,\Phi_4\, \vec q_3 + 1 \,
\end{pmatrix*}
\end{equation}

We will relax the polynomial system of Equations \eqref{second-equ} and replace them with the disjunction of two weaker and simpler equations.  These new constraints no longer involve the variables $\vec p_i$, thus significantly reducing the complexity of our system.

As we shall see below in Lemma \ref{reducetolinear}, the consistency of Equation \eqref{linear-sys} implies that the remaining variables satisfy either one of two additional polynomial constraints.  That is, a solution to Equations \eqref{second-equ} is either a solution of $E$ or a solution of $F$.

The new polynomials $E, F$ and the new solutions introduced through this relaxation do not have an obvious physical or quantum computational interpretation.  Rather, we use them as part of a purely mathematical strategy to establish our desired equality of schemes.

\begin{lemma}\label{reducetolinear}There are polynomials $E,F$ in the entries of $\Phi_{i},\vec q_i$ such that \eqref{linear-sys} being consistent implies $E = 0$ or $F = 0$.
\end{lemma}
\begin{proof}
  Using {\tt Mathematica}, we symbolically compute the order 6 minors of the $6 \times 8$ coefficient matrix of Equation \eqref{linear-sys} and find that they are all zero; thus its rank is always at most 5.  We then symbolically row reduce the following augmented matrix using only moves that are valid whenever $\vec q_4 \neq 0$:
  
 $$ \begin{pmatrix}[cccc|l]
    \vec q_2^{\;\, t} & - \vec q_1^{\;\, t} & 0 & 0 & \vec q_2^{\;t} \,\Phi_1\, \vec q_2 - \vec q_1^{\;t} \,\Phi_2\, \vec q_1 + 1\,   \\
    \vec q_3^{\;\, t} & 0 & - \vec q_1^{\;\, t} & 0 &  \vec q_3^{\;t} \,\Phi_1\, \vec q_3 - \vec q_1^{\;t} \,\Phi_3\, \vec q_1 \,\\
    \vec q_4^{\;\, t} & 0 & 0 & - \vec q_1^{\;\, t}  & \vec q_4^{\;t} \,\Phi_1\, \vec q_4 - \vec q_1^{\;t} \,\Phi_4\, \vec q_1 \,\\
    0 & \vec q_3^{\;\, t} & - \vec q_2^{\;\, t} & 0  &  \vec q_3^{\;t} \,\Phi_2\, \vec q_3 - \vec q_2^{\;t} \,\Phi_3\, \vec q_2 \,\\
    0 & \vec q_4^{\;\, t} & 0 & - \vec q_2^{\;\, t}  & \vec q_4^{\;t} \,\Phi_2\, \vec q_4 - \vec q_2^{\;t} \,\Phi_4\, \vec q_2 \,\\
    0 & 0 & \vec q_4^{\;\, t} & - \vec q_3^{\;\, t}  & \vec q_4^{\;t} \,\Phi_3\, \vec q_4 - \vec q_3^{\;t} \,\Phi_4\, \vec q_3 + 1\\
\end{pmatrix}$$ and find a (rather complicated) rational function $E/F$ in the bottom-right corner.  Here, $F = q_{31} q_{42}-q_{32} q_{41}$.

Fix a choice of values for the variables in the matrix above from $\Z_d$ and assume, given these values, that $F \neq 0$.  From this assumption, we can conclude that $\vec q_4 \neq 0$.  In this case, the result of first substituting the variables for their values and then row reducing the matrix matches the result of symbolically row reducing the matrix and then substituting the variables; this is because every step in our symbolic row reduction is valid when $\vec q_4 \neq 0$.  Thus, the bottom right corner is $E/F$ evaluated at the chosen values.  Since the coefficient  matrix is not full-rank, its bottom row is all zeroes.  So, if the linear system is consistent, $E/F = 0$ implying that $E = 0$.

Thus, consistency of Equation \eqref{linear-sys} implies that $E = 0$ or $F = 0$.
\end{proof}

The {\tt Mathematica} code generating the polynomials $E,F$ is available at \url{https://github.com/ndesilva/semiclifford/}.  These polynomials are explicitly given in Appendix \ref{polyapp}.

\subsection{Schemes of interest}\label{schemesoi}

Here, we define the key schemes over $\Z[1/2]$ we will need to consider in our proof.  We have already defined $T$ whose defining equations are the polynomial systems describing simplified third-level gates (with $\Phi_1 = 0$).  Further, $S$ is defined by adding the semi-Clifford condition of Equation \eqref{semicliffpoly}.  For both $T,S$, we introduce variations for $E$ (resp. $F$) wherein Equations \eqref{second-equ} are replaced with $E = 0$ (resp. $F=0$).  The table below defines schemes via the polynomial generators of their defining ideals.

\begin{table}[H]
\centering
\begin{tabular}{ll}
\textbf{Scheme} &  \textbf{Defining polynomial systems}\\
$\Tr$ &  \eqref{first-equ}, \eqref{second-equ}, \,\,\,\,$\Phi_1 = 0$ \\
$\Sr$ &  \eqref{first-equ}, \eqref{second-equ}, \,\,\,\,$\Phi_1 = 0$, \eqref{semicliffpoly}\\
$\TrE$ &  \eqref{first-equ}, $E=0$, $\Phi_1 = 0$\\
$\TrF$ &  \eqref{first-equ}, $F=0$, $\Phi_1 = 0$\\
$\SrE$ &  \eqref{first-equ}, $E=0$, $\Phi_1 = 0$, \eqref{semicliffpoly}\\
$\SrF$ &  \eqref{first-equ}, $F=0$, $\Phi_1 = 0$, \eqref{semicliffpoly}\\
\end{tabular}
\end{table}
The schemes $T^E$ and $T^F$ are defined by a weakened polynomial system and may have extraneous components in addition to those of $T$ as we will explain later.
Since a solution to Equations \eqref{second-equ} is either a solution of $E = 0$ or a solution of $F = 0$:
\begin{align}
    \Tr\zd &\subseteq \tTrE\zd \cup \tTrF\zd \label{TEF}.
\end{align}

We can now formally frame our proof strategy in terms of the algebraic sets of $\Z_d$-rational points of our above-defined schemes of interest. 
\begin{lemma}\label{mainthmlemma}
If, for any odd prime dimension $d \in \N$,  $\Tr(\Z_d) = \Sr(\Z_d)$ then Theorem \ref{mainthm} holds.
\end{lemma}

\begin{proof}
Let $d$ be an odd prime and $G$ be a two-qudit third-level gate.  By Theorem \ref{diagsymp} there exists a Clifford gate $C_1 \in \cC_2^2$ such that $C_1 G C_1^*$ is a simplified third-level gate.   By Theorem \ref{phireduction}, we can find a Clifford gate $C_2 \in \cC_2^2$ such that $G' = C_1 G C_1^* C_2$ is simplified and the tuple of elements of $\Z_d$ that describe the conjugate tuple of $G'$ is contained in $\Tr(\Z_d)$.  Therefore, it is contained in $\Sr(\Z_d)$ and $G'$ is semi-Clifford.  By Lemma \ref{cliffmultsemi}, $G$ is semi-Clifford.  
\end{proof}

\subsection{Proof of main theorem}\label{proofmain}

\mainthm*

We are now ready to analyse our schemes of interest in order to establish our main theorem.  First, we will establish three lemmas concerning $\TrE$ and $\SrE$; we will then perform the same analyses with $\TrF$ and $\SrF$ and achieve the same results \emph{mutatis mutandis}.  

The thrust of the following lemmas will be to algebraically study and compare the components of $\TrE$, $\SrE$ and  $\TrF$, $\SrF$.  We will find that $\TrE$ decomposes into five components: three of which compose $\SrE$ whereas the last two are simply byproducts of relaxing Equations \eqref{second-equ} and do not describe genuine third-level gates.  The same conclusions hold for $\TrF$, $\SrF$.  We will then conclude that every $\Z_d$-rational point of $T$ is also a point of either $\SrE$ or $\SrF$ and therefore, represents a semi-Clifford gate.

The {\tt Magma} code for these calculations is available at \url{https://github.com/ndesilva/semiclifford/}.

\begin{remark}
    Recall that the defining equations of $T^E, T^F, S^E, S^F$ do not involve the variables $\vec{p}_i$, which is what renders the computational verification of the following Lemmas \ref{ST-decompose} and \ref{ST-match} feasible.  However, in our computations, we retain these variables in defining our affine spaces.  Doing so adds minimal computational overhead and simplifies the arguments of Lemma \ref{disjoint}.
\end{remark}

\subsubsection{Lemmata}

\begin{lemma}
\label{ST-decompose}
The scheme $\TrE$ over $\Z[1/2]$ decomposes into $5$ closed subschemes over $\Z[1/2]$
 \begin{equation}
 \label{S-decompose}
   \TrE = \TrE_1 \cup \TrE_2 \cup \TrE_3 \cup \TrE_4 \cup \TrE_5
 \end{equation}
 of dimension $16, 16, 17, 17, 17$ over $\Q$, respectively, and the scheme $\SrE$ over $\Z[1/2]$ decomposes into $3$ closed subschemes over $\Z[1/2]$
 \begin{equation}
 \label{T-decompose}
   \SrE = \SrE_1 \cup \SrE_2 \cup \SrE_3
\end{equation}
of dimension $16, 16, 17$ over $\Q$, respectively. Furthermore, we have that
\begin{align}
     \TrE(\Z_d) & =  \TrE_1(\Z_d) \cup \TrE_2(\Z_d) \cup \TrE_3(\Z_d) \cup \TrE_4(\Z_d) \cup \TrE_5(\Z_d), \label{TEdecomp}\\
     \SrE(\Z_d) & = \SrE_1(\Z_d) \cup \SrE_2(\Z_d) \cup \SrE_3(\Z_d). \label{SEdecomp}
\end{align}
\end{lemma}
\begin{proof}
We achieve this by using the method described at the end of Section~\ref{decomposition} with the scheme $X = \TrE, \SrE$ over $\Z$ (the defining equations of $\TrE, \SrE$ have coefficients in $\Z$ so they define a scheme over $\Z$).

That is, we use the {\tt Magma} command {\tt ProbableRadicalDecomposition} which takes as input the polynomials that (generate the ideal that) defines the scheme $\TrE$ and returns as output the lists of polynomial generators (of the ideals that) define the components $\TrE_1, ..., \TrE_5$.  The decomposition is performed over $\Q$ and the results are verified to be valid over $\Z[1/2]$.

We perform a similar calculation for $\SrE$ to obtain descriptions of the components $\SrE_1, ..., \SrE_3$ as three radical ideals described concretely as lists of polynomial generators.
\end{proof}

Crucially, the first three components of $\TrE$ and $\SrE$ are the same.
\begin{lemma}
\label{ST-match}
The schemes over $\Z[1/2]$,  $\SrE_1, \SrE_2, \SrE_3$ of $\SrE$ match the components $\TrE_1, \TrE_2, \TrE_3$ of $\TrE$, respectively, that is: $\SrE_i = \TrE_i$ as schemes over $\Z[1/2]$ for all $i = 1, 2, 3$.
\end{lemma}
\begin{proof}
We will establish this lemma by extracting, from the {\tt Magma} computations that establish Lemma \ref{ST-decompose}, the polynomials that define the components $\TrE_1$ and $\SrE_1$ and showing that both ideals of polynomial define the same component.  We repeat this for $\TrE_2$, $\SrE_2$ and $\TrE_3$, $\SrE_3$.

We achieve this by using noting that the generators for $I(\SrE_i)$ and $I(\TrE_i)$ have coefficients in $\Z[1/2]$ and we check in {\tt Magma} that each generator of $I(\SrE_i)$ is a polynomial linear combination of the generators of $I(\TrE_i)$ and vice versa.  By \emph{polynomial linear combination}, we mean a linear combination of the generators whose coefficients are polynomials in the variables given by the entries of $\vec{q}_i$ and $\Phi_i$, for $1 \le i \le 4$, with coefficients in $\Z[1/2]$.
\end{proof}

Thus, we can conclude that:
\begin{equation}\label{TS123}
    \TrE_i\zd = \SrE_i\zd \text{ for } i=1,2,3.
\end{equation}

Next, we will show that the $\Z_d$-rational points of the last two components $\TrE_4, \TrE_5$ of $\Tr^E$ do not correspond to valid third-level gates (which is the meaning of \eqref{extraneous} below). 


\begin{lemma}
\label{disjoint}
The components $\TrE_4, \TrE_5$ are extraneous in the sense that \begin{equation}
\tTrE_4(\Z_d) \cap \Tr(\Z_d) = \tTrE_5(\Z_d) \cap \Tr(\Z_d) = \emptyset.\label{extraneous}
\end{equation}
\end{lemma}
\begin{proof}
We will establish this lemma by extracting, from the {\tt Magma} computations that establish Lemma \ref{ST-decompose}, the polynomials that define the components $\TrE_4, \TrE_5$.  We will show that they are inconsistent with Equations 
\eqref{second-equ} and, therefore, the points of these components do not describe valid third-level gates.  

The $\vec p_i, \vec q_i$ below are not derived from gates, but rather are simply variables used in the polynomials that define the components of our schemes $\Tr, \tTrE_4, \tTrE_5$.

The equations for $\tTrE_4$ are 
\begin{equation}
  \vec q_1 = \ldots = \vec q_4 = 0.
\end{equation}
However, for $(i,j) \in \left\{ (1,2), (3,4) \right\}$, we obtain a contradiction to the equation in \eqref{second-equ}.

The equations for $\tTrE_5$ include the equations
\begin{align}
  \vec q_1 & = 0 \\
  \lambda_3 \vec q_3 & = \mu_2 \vec q_2 \\
  \lambda_4 \vec q_4 & = \nu_2 \vec q_2
\end{align}
for some scalars $\lambda_3, \mu_2$; $\lambda_4, \nu_2$, with each pair not both zero.

If either $\lambda_3 = 0$ or $\lambda_4 = 0$, then $\vec q_2 = 0$, but then for $(i,j) = (1,2)$, the left hand side of \eqref{second-equ} is $0$, while the right hand side of \eqref{second-equ} is $1$; a contradiction.

Assume $\lambda_3$ and $\lambda_4$ are both nonzero. Then $\vec q_3$ and $\vec q_4$ are scalar multiples of $\vec q_2$. The equations in \eqref{second-equ} for $(i,j) = (1,2), (1,3), (1,4)$ read as
\begin{align}
    \vec p_1 \cdot \vec q_2 & = 1 \\
    \vec p_1 \cdot \vec q_3 & = 0 \\
    \vec p_1 \cdot \vec q_4 & = 0.
\end{align}
The last two equations imply that $\vec q_3 = \vec q_4 = 0$. But then \eqref{second-equ} for $(i,j) = (3,4)$ has left hand side being $0$ but right hand side being $1$ which is a contradiction.

In summary, we deduce that $\tTrE_4(\Z_d) \cap \Tr(\Z_d) = \tTrE_5(\Z_d) \cap \Tr(\Z_d) = \emptyset$.
\end{proof}

Combining \eqref{TEdecomp}, \eqref{SEdecomp}, \eqref{TS123} and \eqref{extraneous}, we find that:
\begin{equation}
(\tTrE\zd \cap \Tr\zd)  = (\tSrE\zd \cap \Tr\zd).\label{TSE}
\end{equation}
Repeating the calculations of the above two lemmas with $\TrF,\SrF$, we find that:
\begin{equation}
(\tTrF\zd \cap \Tr\zd)  = (\tSrF\zd \cap \Tr\zd)\label{TSF}
\end{equation}
Finally, we can use the above two lemmas to conclude that third-level gates are semi-Clifford.

\subsubsection{Main theorem}

Here, we formalise how to conclude our main theorem from the fact that $\TrE$ (resp. $\TrF$) and  $\SrE$ (resp. $\SrF$) are the same up to components that do not describe valid third-level gates.

\begin{theorem}\label{ST}
  For any odd prime dimension $d \in \N$,  $\Tr(\Z_d) = \Sr(\Z_d)$.
\end{theorem}
\begin{proof}
Clearly, $\Sr(\Z_d) \subseteq \Tr(\Z_d)$. We also have that
\begin{align}
    \Tr(\Z_d) & \subseteq (\tTrE\zd \cup \tTrF\zd) \cap \Tr\zd \label{eq:1} \\
    & = (\tTrE\zd \cap \Tr\zd) \cup (\tTrF\zd \cap \Tr\zd) \\
    & = (\tSrE\zd \cap \Tr\zd) \cup (\tSrF\zd \cap \Tr\zd) \label{eq:2}\\
    & = (\tSrE\zd \cup \tSrF\zd) \cap \Tr\zd \\
    & \subseteq \Sr(\Z_d) \label{eq:3}.
\end{align}

Here, Equation (\ref{eq:1}) follows from \eqref{TEF}: $\Tr(\Z_d) \subseteq \tTrE\zd \cup \tTrF\zd$.  Equation (\ref{eq:2}) follows from \eqref{TSE} and \eqref{TSF}.  Finally, Equation (\ref{eq:3}) follows from the facts that every element of $\Tr\zd$ satisfies Equations \eqref{first-equ}, \eqref{second-equ} and $\Phi_1 = 0$ and that every element of $\tSrE\zd,\tSrF\zd$ satisfies the semi-Clifford condition of Equation \eqref{semicliffpoly}.\end{proof}

Using Lemma \ref{mainthmlemma}, Theorem \ref{ST} entails Theorem \ref{mainthm}.

\section{Conclusions}

A natural follow-up question to our work is to generalise our result to higher levels of the Clifford hierarchy.  We can also consider extending our techniques to generalise the counterexamples of Zeng-Chen-Chuang \cite{zeng2008semi} and Gottesman–Mochon to higher dimensions; that is, find examples of $n$-qudit $k$-th level gates that are not semi-Clifford when $n >2, k>3$ or $n>3, k =3$.

This work significantly advances the program of classifying gates of the Clifford hierarchy and semi-Clifford gates.  Deeper mathematical understanding of the Clifford hierarchy and semi-Clifford gates will lead to more efficient circuit and gate synthesis.  It further bolsters the viability of qudit-based fault-tolerant universal quantum computers by providing complete sets of efficient gate teleportation protocols.  This is practically important as qudit magic state distillation has been proposed as a significantly more efficient alternative to the qubit case \cite{quditmsd}.  This, and other advantages of qudits, are driving current experimental research \cite{chi2022programmable,chizzini2022molecular,karacsony2023efficient,low2020practical,ringbauer2022universal,seifert2022time,wang2018proof}.

By clarifying the algebraic structure of magic states, we contribute to the understanding of resources that power quantum computation.  Howard et al. \cite{howard2014contextuality} showed that, for odd prime $d$, all distillable one-qudit magic states, once tensored with any other state, exhibit contextuality with respect to two-qudit stabiliser measurements.  The second author \cite{de2018logical} showed that, for odd prime $d$ with $d \not\equiv 1 \textrm{ (mod 3)}$, magic states for diagonal third-level two-qudit gates exhibit the paradoxical form of strong contextuality.  By showing that the most powerful magic states witness the logically strongest form of contextuality, the latter result bolsters a refinement of the resource theory of contextuality that emphasises the computational power of logical paradoxes.  Theorem \ref{mainthm} strengthens it by establishing that these magic states are sufficient for implementing any third-level two-qudit gate.

The abstract mathematical techniques developed to solve our problem are widely applicable to many more problems within quantum information.  We give a blueprint for solving any problem that can be recast in terms of the equivalence of solution sets of polynomial equations over $\Z_d$, for all odd prime $d$.  This is a potentially very broad class of problems given that the dominant stabiliser formalism for quantum error correction is based on the standard representation of the Heisenberg-Weyl group over $\Z_d$.

\section{Acknowledgments}
IC was supported by the Natural Sciences and Engineering Research Council of Canada (NSERC): Discovery Grant RGPIN-2023-03457.  ND acknowledges support from the Canada Research Chair program, NSERC Discovery Grant RGPIN-2022-03103, the NSERC-European Commission project FoQaCiA, and the Faculty of Science of Simon Fraser University.

\section{Declarations}
The authors have no relevant financial or nonfinancial interests or competing interests to declare that are relevant to the content of this article.

\bibliographystyle{abbrv}
{\footnotesize
\bibliography{cliff}}

\appendix
\section{The polynomials $E,F$}\label{polyapp}

Here, we explicitly give the polynomials $E,F$ of Lemma \ref{reducetolinear}.  The polynomial $F$ is $q_{31} q_{42}-q_{32} q_{41}$ and the polynomial $E$ is:\\
\noindent $q_{11} q_{22}-q_{12} q_{21} + q_{31} q_{42}-q_{32} q_{41} + \\
\Phi_{11} \left(-q_{32} q_{41} q_{21}^2+q_{31} q_{42} q_{21}^2+q_{32} q_{41}^2 q_{21}-q_{31}^2 q_{42} q_{21}-q_{22} q_{31} q_{41}^2+q_{22} q_{31}^2 q_{41}\right)+\\
\Phi_{12} \left(q_{31} q_{42} q_{22}^2-q_{31} q_{42}^2 q_{22}+q_{32}^2 q_{41} q_{22}+q_{21} q_{32} q_{42}^2-q_{22}^2 q_{32} q_{41}-q_{21} q_{32}^2 q_{42}\right)+\\
\Phi_{13} \left(-q_{21} q_{22} q_{32} q_{41}+q_{22} q_{31} q_{32} q_{41}-q_{22} q_{31} q_{42} q_{41}+q_{21} q_{32} q_{42} q_{41}+q_{21} q_{22} q_{31} q_{42}-q_{21} q_{31} q_{32} q_{42}\right)+\\
\Phi_{21} \left(q_{32} q_{41} q_{11}^2-q_{31} q_{42} q_{11}^2-q_{32} q_{41}^2 q_{11}+q_{31}^2 q_{42} q_{11}+q_{12} q_{31} q_{41}^2-q_{12} q_{31}^2 q_{41}\right)+\\
\Phi_{22} \left(q_{32} q_{41} q_{12}^2-q_{31} q_{42} q_{12}^2+q_{31} q_{42}^2 q_{12}-q_{32}^2 q_{41} q_{12}-q_{11} q_{32} q_{42}^2+q_{11} q_{32}^2 q_{42}\right)+\\
\Phi_{23} \left(q_{11} q_{12} q_{32} q_{41}-q_{12} q_{31} q_{32} q_{41}+q_{12} q_{31} q_{42} q_{41}-q_{11} q_{32} q_{42} q_{41}-q_{11} q_{12} q_{31} q_{42}+q_{11} q_{31} q_{32} q_{42}\right)+\\
\Phi_{31} \left(-q_{22} q_{41} q_{11}^2+q_{21} q_{42} q_{11}^2+q_{22} q_{41}^2 q_{11}-q_{21}^2 q_{42} q_{11}-q_{12} q_{21} q_{41}^2+q_{12} q_{21}^2 q_{41}\right)+\\
\Phi_{32} \left(q_{21} q_{42} q_{12}^2-q_{21} q_{42}^2 q_{12}+q_{22}^2 q_{41} q_{12}+q_{11} q_{22} q_{42}^2-q_{12}^2 q_{22} q_{41}-q_{11} q_{22}^2 q_{42}\right)+\\
\Phi_{33} \left(-q_{11} q_{12} q_{22} q_{41}+q_{12} q_{21} q_{22} q_{41}-q_{12} q_{21} q_{42} q_{41}+q_{11} q_{22} q_{42} q_{41}+q_{11} q_{12} q_{21} q_{42}-q_{11} q_{21} q_{22} q_{42}\right)+\\
\Phi_{41} \left(q_{22} q_{31} q_{11}^2-q_{21} q_{32} q_{11}^2-q_{22} q_{31}^2 q_{11}+q_{21}^2 q_{32} q_{11}+q_{12} q_{21} q_{31}^2-q_{12} q_{21}^2 q_{31}\right)+\\
\Phi_{42} \left(q_{22} q_{31} q_{12}^2-q_{21} q_{32} q_{12}^2+q_{21} q_{32}^2 q_{12}-q_{22}^2 q_{31} q_{12}-q_{11} q_{22} q_{32}^2+q_{11} q_{22}^2 q_{32}\right)+\\
\Phi_{43} \left(q_{11} q_{12} q_{22} q_{31}-q_{12} q_{21} q_{22} q_{31}+q_{12} q_{21} q_{32} q_{31}-q_{11} q_{22} q_{32} q_{31}-q_{11} q_{12} q_{21} q_{32}+q_{11} q_{21} q_{22} q_{32}\right).
$

\end{document}